\documentclass{article}

\usepackage[preprint]{neurips_2024}

\usepackage[utf8]{inputenc} 
\usepackage[T1]{fontenc}    
\usepackage{hyperref}       
\usepackage{url}            
\usepackage{booktabs}       
\usepackage{amsfonts}       
\usepackage{nicefrac}       
\usepackage{microtype}      
\usepackage{xcolor}         
\usepackage{comment}
\usepackage{natbib}
\title{MERaLiON-SER: Robust Speech Emotion Recognition Model for English and SEA Languages}

\author{
  MERaLiON Team \\
  Institute for Infocomm Research (I$^{2}$R), A*STAR, Singapore \\
  Corresponding Author: Hardik B. Sailor\\
  \texttt{sailorhb@a-star.edu.sg} \\
}
\usepackage{booktabs,siunitx,adjustbox}
\usepackage{makecell} 
\begin{document}

\maketitle

\begin{abstract}
We present \textbf{MERaLiON-SER}\footnote{Huggingface model and demo is available here: https://huggingface.co/MERaLiON/MERaLiON-SER-v1}, a robust speech emotion recognition model designed for English and Southeast Asian languages. The model is trained using a hybrid objective combining weighted categorical cross-entropy and Concordance Correlation Coefficient (CCC) losses for joint discrete and dimensional emotion modelling. This dual approach enables the model to capture both the distinct categories of emotion (like 'happy' or 'angry') and the fine-grained, such as arousal (intensity), valence (positivity/negativity), and dominance (sense of control), leading to a more comprehensive and robust representation of human affect.
Extensive evaluations across multilingual Singaporean languages (English, Chinese, Malay, and Tamil ) and other public benchmarks show that MERaLiON-SER consistently surpasses both open-source speech encoders and large Audio-LLMs. These results underscore the importance of specialised speech-only models for accurate paralinguistic understanding and cross-lingual generalisation. Furthermore, the proposed framework provides a foundation for integrating emotion-aware perception into future \textit{agentic audio systems}, enabling more empathetic and contextually adaptive multimodal reasoning.
\end{abstract}

\section{Introduction}

Human speech is a rich medium not only for semantic communication, but also for conveying emotional and paralinguistic information such as affective states, intention, and interpersonal stance. Recognising emotions from speech has become a foundational task in affective computing and human-machine interaction: accurate speech emotion recognition (SER) enables applications ranging from virtual assistants that adapt to user mood, to mental-health monitoring, to empathetic conversational agents in service and robotics domains.  

Yet despite impressive progress in speech processing and representation learning, SER remains significantly more challenging than many seemingly ``solved" tasks in speech and language.  First, emotional expression in speech is inherently ambiguous and subjective: different listeners often disagree on the label of a given utterance, mixed emotions may co-occur, and prosodic cues can vary wildly across speakers, languages and cultures. For example, work on annotation ambiguity shows that labelling disagreements among raters are common and that collapsing those disagreements into a single “majority” label may discard useful signal.  

Second, SER suffers from annotation scarcity and imbalance: many datasets are small, acted rather than spontaneous, and heavily biased toward a handful of well-studied languages (e.g., English). Under-resourced languages or code-mixed scenarios (typical in multilingual regions) are far less represented.  

Third, cross-domain and cross-corpus transfers remain problematic: models trained on one dataset (speakers, recording conditions, culture) often degrade drastically when applied to a different domain. Domain shifts include language, speaker demographics, recording channel or environment noise.

Fourth, although large self-supervised and multimodal models (e.g., speech encoders or audio-language models) provide powerful acoustic and semantic representations, they often prioritize linguistic content (ASR or transcription) over paralinguistic cues \citep{chen2025audio}. As a result, even state-of-the-art audio-LLMs may under-perform when tasked with fine-grained emotion reasoning in multilingual or culturally diverse scenarios.  

In multilingual and culturally diverse regions (such as Southeast Asia), emotional expression can be shaped by language mixing, prosodic variation, and code-switching. These phenomena demand models that capture both linguistic and paralinguistic signals, adapt across languages and speakers, and remain robust to annotation and domain shift.  

In this work, we present \textbf{MERaLiON-SER}, a robust, parameter-efficient speech emotion recognition model tailored for English and Southeast Asian languages. As the model was developed under the MERaLiON project, it is accordingly designated as MERaLiON-SER. 
Unlike general audio-language models, which are optimised primarily for speech-to-text or audio-text tasks, MERaLiON-SER emphasises paralinguistic specialisation for learning emotion intent directly from speech, independent of transcription. Our experiments show strong generalisation across Singaporean multilingual code-mixed speech, Southeast Asian languages, and public benchmark datasets. This design positions MERaLiON-SER as a key building block towards emotionally intelligent, multilingual and agentic audio systems.

\section{Model description}
\label{model_description}
The proposed architecture employs the Whisper-Medium encoder as the foundational backbone for multilingual acoustic feature extraction. Building upon this encoder, a custom downstream network is introduced, comprising attention-based pooling layers followed by modified Emphasized Channel Attention, Propagation and Aggregation in Time Delay Neural Network (ECAPA-TDNN) modules to capture both temporal dynamics and speaker-invariant paralinguistic features \citep{ECAPA}. Compared to the original ECAPA-TDNN model, we replaced all BatchNorm layers with GroupNorm layers. The model adopts a multi-head output design to jointly address categorical and dimensional emotion recognition tasks. Specifically, the categorical head branch includes a softmax layer for seven discrete emotions, while the dimensional head branch includes a sigmoid layer to estimate three continuous dimension scores. \textbf{This model supports seven emotion classes, namely, Neutral, Happy, Sad, Angry, Surprised, Fearful, Disgusted and  three dimensional emotion scores for arousal, valence, dominance.} This joint learning approach facilitates fine-grained and segmental prosody rich representation of emotional expressions.

\begin{figure}
    \centering
    \includegraphics[width=0.5\linewidth]{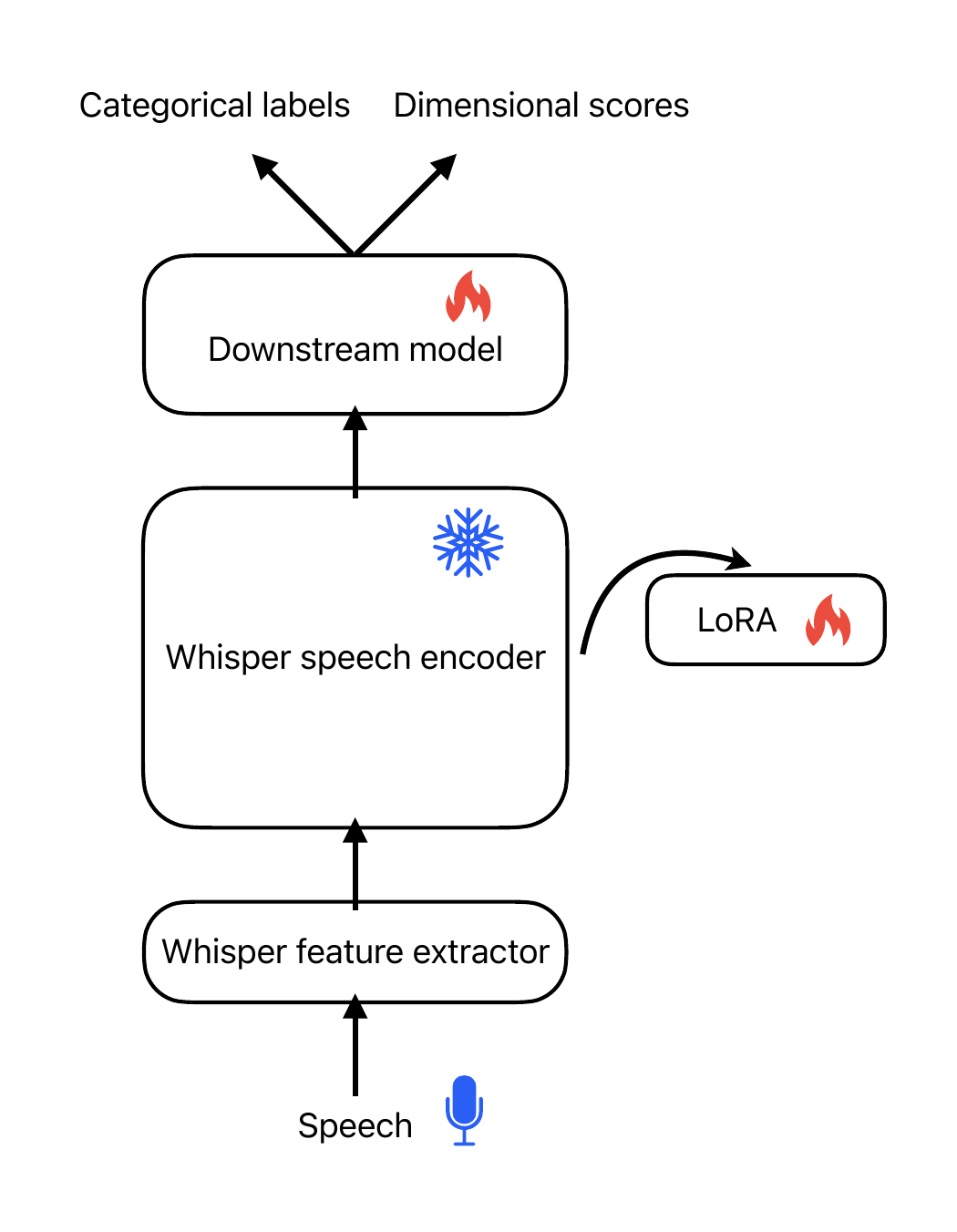}
    \caption{Block diagram of proposed model}
    \label{fig:serfig}
\end{figure}

To mitigate overfitting and computational overhead during training, the Whisper encoder parameters are kept frozen, while Low-Rank Adaptation (LoRA) adapters are integrated into the attention layers (key, query, and value). This training approach enable efficient task-specific fine-tuning \citep{hu2022lora}. Furthermore, to capture emotional cues at multiple temporal resolutions, we introduce novel multiscale and hierarchical attention pooling techniques that enable the model to effectively combine short and long-term emotion cues. The overall block diagram of the model is shown in Figure \ref{fig:serfig}. The downstream model includes ECAPA-TDNN pooling layers, and task-specific output heads.

Overall, the model is designed with an emphasis on parameter-efficient multilingual affective modeling and temporal abstraction, enabling the extraction of expressive emotional representations without extensive retraining or large-scale parameter updates. This design philosophy balances computational efficiency, generalization capability, and cross-lingual transferability, making the model suitable for scalable deployment in diverse speech-based affective computing applications


\section{Training Details}

\subsection{Loss Functions}
The overall objective combined categorical and dimensional emotion losses. The categorical branch optimized a weighted cross-entropy loss with label smoothing = 0.1, while the dimensional branch employed the Concordance Correlation Coefficient (CCC) loss. The total loss was a weighted sum of these components with coefficients \(\lambda_{\text{cat}} = 1.0\) and \(\lambda_{\text{dim}} = 0.5\).

The loss function was a weighted sum of the following components:

\textbf{Weighted Cross-Entropy Loss (CE)} for the categorical emotion classification:
\begin{equation}
\mathcal{L}_{\text{CE}} = - \sum_{i=1}^{C} w_i \cdot y_i \log(\hat{y_i}),
\end{equation}
where \(C\) is the number of emotion classes, \(y_i\) is the ground truth for the \(i\)-th class, \(\hat{y_i}\) is the predicted probability, and \(w_i\) is the weight assigned to the \(i\)-th class to handle class imbalance.

\textbf{Concordance Correlation Coefficient (CCC) Loss} for the dimensional emotion regression:
\begin{equation}
\mathcal{L}_{\text{CCC}} = 1 - \frac{2 \cdot \text{cov}(y, \hat{y})}{\text{var}(y) + \text{var}(\hat{y}) + (\mu_y - \mu_{\hat{y}})^2},
\end{equation}
where \(y\) and \(\hat{y}\) are the true and predicted dimensional values, respectively, \(\mu_y\) and \(\mu_{\hat{y}}\) are the means of the true and predicted values, and \(\text{cov}(y, \hat{y})\) and \(\text{var}(y)\), \(\text{var}(\hat{y})\) are the covariance and variance terms.

The combined loss function is a weighted sum of the above components:
\begin{equation}
\mathcal{L} = \lambda_{\text{cat}} \mathcal{L}_{\text{CE}} + \lambda_{\text{dim}} \mathcal{L}_{\text{CCC}},
\end{equation}
where \(\lambda_{\text{cat}}\) and \(\lambda_{\text{dim}}\) are the scaling coefficients for the categorical and dimensional, respectively.

\subsection{Training Configuration}
Optimization was carried out using two parameter groups: a low learning rate for the frozen backbone with LoRA adapters (learning rate=5e-5, weight decay = 4e-5) and a higher rate for the downstream network (learning rate=6e-4, weight decay = 8e-5). A cosine annealing scheduler with a 0.08 warm-up ratio was employed to gradually adjust the learning rate during training. To improve model generalization, label smoothing with an $\epsilon$ of 0.1 was employed during training. The model was trained for 15 epochs with a batch size of 32 on single node with 8 Nvidia H100 GPUs. We used development set categorical loss as early stopping criteria. 

During training, extensive data augmentations were applied to ensure robustness to environmental and speaker variations. MixUp was performed with a probability of 0.5 and mixing coefficient \(\alpha = 0.3\) \citep{zhang2017mixup}. Additional augmentations include additive noise drawn from the MUSAN corpora \citep{snyder2015musan} and speed perturbations with factors of 0.9 and 1.1, respectively. 

\section{Datasets}
\subsection{Training datasets}
The MERLION-SER was trained using a combination of proprietary labeled, pseudo-labeled, and open-source datasets that allows usage in commercial model building.
\subsubsection{SG training dataset}
Two training datasets were used in this model that include SG languages: the SG-ECMT train set and the SGTV dataset.

\textbf{SG-ECMT Train Set}: The SG-ECMT train set consists of speech data in English, Malay, Chinese, and Tamil, the four major languages spoken in Singapore. This data is derived from our proprietary unlabeled raw speech corpora, containing 10–30 seconds segments across the following seven emotion categories: neutral, angry, disgusted, fearful, happy, sad, and surprised.

The raw data was processed and filtered following the pipeline described in \citep{cpqa_interspeech}. Specifically, emotion labels were estimated every 4 seconds with a 2 seconds overlap using the emotion2vec pipeline \citep{emotion2vec}, which classifies speech into nine categories: the seven target emotions listed above, plus \textit{other} and \textit{unknown}.

To ensure high-quality labeling, we retained only the segments where both the emotion2vec+ seed and emotion2vec+ base models produced identical predictions among the six target emotions (angry, disgusted, fearful, happy, sad, surprised). Speech segments without consistent emotion predictions were labeled as neutral. Speech samples were then selected based on the number of emotional segment occurrences to ensure reliable emotion representation within each utterance \citep{wang2025benchmarking}.

The resulting SG-ECMT dataset comprises 27,458, 14,212, 14,370, and 10,169 samples for the English, Chinese, Malay, and Tamil training sets, respectively. 
This pseudo-labeled dataset was subsequently used to generate a second set of pseudo labels using our MERaLiON-SER model in a two-pass labeling process.

\textbf{SGTV Dataset:} The SGTV dataset is an internally created, human-labeled emotional speech dataset. This data consists of approximately 117,000 speech samples (about 120 hours) collected from Singaporean TV shows and movies. Data primarily include English, Mandarin Chinese and code-mixed utterances between the two languages. Emotion labels were manually annotated by trained human annotators and include the seven categories from the SG-ECMT train set.

\subsubsection{Open source training set}
We trained the model using publicly available open-source datasets. This includes CREMA-D \citep{cao2014crema}, M3ED \citep{huang2021m3ed}, ESD \citep{zhou2018esd}, and MELD \citep{meld}.

\subsection{Evaluation dataset}
The model's performance was rigorously evaluated on both manually validated Singapore language sets and public multilingual benchmarks.

\subsubsection{Manually curated SG evaluation set}
We curated the SG-ECMT evaluation set following the same procedure as the SG-ECMT train set. It consists of 1,880 speech samples, each ranging from 10–30 seconds in duration. Initially, we selected 80 samples per emotion category per language for English, Chinese, Malay, and Tamil.

Subsequently, three native-speaking human validaters for each language manually reviewed and corrected the automatically generated emotion labels. Then a majority-vote scheme was applied to finalize the labels.

In total, the resulting evaluation sets comprise 466, 466, 479, and 469 samples for English, Chinese, Malay, and Tamil, respectively.

\subsubsection{Public evaluation set}
We also used popular public datasets for English, Chinese, and Indonesia. For English, we used MSP-podcast test1 \citep{busso2017msp}, IEMOCAP five test folds \citep{iemocap}, and MELD. We also included M3ED for Chinese and IndoWaveSentiment for Indonesia.

\section{Evaluation setup} 
For evaluation, we averaged the predictions from the top four model checkpoints, selected based on the lowest categorical loss on the development set, to ensure robust and stable performance. We benchmarked our proposed model against both open-source and closed-source systems covering a range of architectures and modalities.

\paragraph{Open-source Speech-only Models:}
We included three variants of \textbf{emotion2vec}--\textit{Large}, \textit{Base}, and \textit{Seed} that differ in model capacity and pretraining data. These models represent strong speech-only baselines for emotion recognition.

\paragraph{Open-source Audio LLMs:}
To compare with recent large-scale multimodal speech models optimized for Southeast Asian (SEA) languages, we evaluated \textbf{MERaLiON-10B}~\citep{meralion} and \textbf{SeaLLMs-Audio-7B}~\citep{SeaLLMs-Audio}. Both models incorporate speech-text alignment during finetuning and demonstrate strong cross-lingual generalization.

\paragraph{Closed-source Multimodal Models:}
For completeness, we further compared our model with two state-of-the-art proprietary systems: \textbf{GPT-4o-Audio}~\citep{gpt4} and \textbf{Gemini-2.5-Flash}~\citep{gemini}---which are capable of joint reasoning across audio and text modalities.

\paragraph{Prompt and Inference Configuration:}
A uniform prompt was used for fair evaluation across all LLM-based models:
\begin{quote}
\texttt{PROMPT = "Determine the speaker's emotion in the given audio. Reply with a single label from: Neutral, Happy, Sad, Angry, Fearful, Disgusted, Surprised."}
\end{quote}

\begin{table}[h]
\centering
\caption{Summary of baseline and comparison models used for evaluation.}
\label{tab:models}
\begin{tabular}{lcccc}
\toprule
\textbf{Model} & \textbf{Type} & \textbf{Parameters} & \textbf{Availability} & \textbf{License} \\
\midrule
\multicolumn{5}{l}{\textit{Open-source Speech-only Models}} \\
emotion2vec-large & Speech-only Encoder & 300M & Open-source & Not mentioned*\\
emotion2vec-base & Speech-only Encoder & 90M & Open-source & Not mentioned*\\
emotion2vec-seed & Speech-only Encoder & 90M & Open-source & Not mentioned*\\
MERaLiON-SER & Speech-only Encoder & 309M & Open-source & MIT \\
\midrule
\multicolumn{5}{l}{\textit{Open-source Audio LLMs (SEA-optimized)}} \\
MERaLiON-10B & Audio LLM & 10B & Open-source & MIT \\
SeaLLMs-Audio-7B & Audio LLM & 7B & Open-source & Not mentioned\\
\midrule
\multicolumn{5}{l}{\textit{Closed-source Multimodal Models}} \\
GPT-4o-Audio & Multimodal LLM & -- & Closed-source & Paid API \\
Gemini-2.5-Flash & Multimodal LLM & -- & Closed-source & Paid API \\
\bottomrule
\end{tabular}
*All emotion2vec models include several datasets that restrict their commercial usage. 
\end{table}

\section{Results and Discussion}
\label{sec:results}
In SER task, class imbalance is a persistent challenge, as certain emotions such as neutral or happy typically dominate spontaneous datasets, while others like fear or disgust occur infrequently. Conventional performance metrics, such as weighted accuracy or traditional accuracy tend to be dominated by these majority classes, often masking a model’s poor discrimination of minority emotions. To provide a more balanced evaluation, researchers in the affective computing community commonly use Unweighted Average Recall (UAR) also known as Balanced Accuracy as the principal metric. Unlike weighted accuracy which aggregates correct predictions proportional to class distribution—UAR offers a class-independent assessment that more accurately reflects a model’s generalization capability across diverse emotional states.

We have evaluated MERaLiON-SER-v1 performance on evaluation set of SG-ECMT dataset. The SG-ECMT dataset for Singapore languages contains fine-grained labels at every two seconds and merged nearby segments to create a course level segments with maximum of 15 second duration. We have also added performance of primary 4 classes in emotion literature: Neutral, Angry, Sad, and Happy.
\subsection{Evaluation on Singapore Languages}

\begin{figure}
    \centering
    \includegraphics[width=1.0\linewidth]{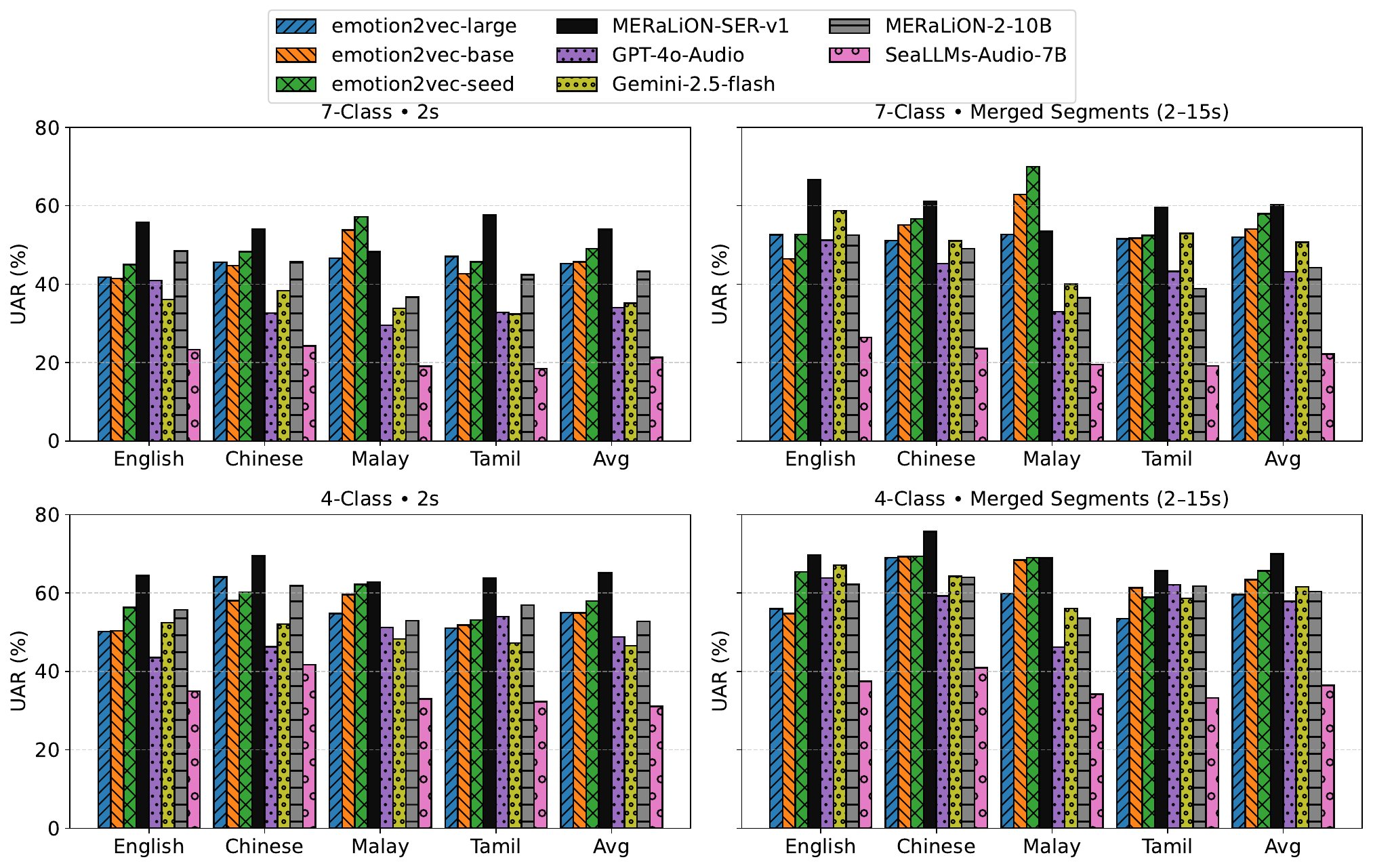}
    \caption{Results for Singapore languages.}
    \label{fig:sg_uar_4plots}
\end{figure}

Figure~\ref{fig:sg_uar_4plots} presents the UAR across four Singapore languages—English, Chinese, Malay, and Tamil—under both 7-class and 4-class emotion settings, evaluated using fine-grained 2-second segments and merged segments ranging from 2--15 seconds.

Overall, \textbf{MERaLiON-SER-v1} achieves the highest average UAR across all configurations, outperforming both open-source speech encoders (\textit{emotion2vec-large, base, seed}) and multimodal LLMs (\textit{GPT-4o-Audio, Gemini-2.5-flash, MERaLiON-2-10B, SeaLLMs-Audio-7B}). 
Specifically, MERaLiON-SER-v1 reaches 53.9\% (7-class, 2s), 60.2\% (7-class merged), 65.1\% (4-class, 2s), and 70.0\% (4-class merged), 
surpassing the best open-source speech encoder (\textit{emotion2vec-seed}) by +4.9, +2.3, +7.1, and +4.3 absolute UAR points respectively.

\paragraph{Speech-only vs. multimodal LLMs:}
Among the speech-only models, the performance ranking is consistent across all settings:
\textbf{MERaLiON-SER-v1} $>$ \textit{emotion2vec-seed} $>$ \textit{emotion2vec-base} $\approx$ \textit{emotion2vec-large}.
Audio-LLMs, though trained with large-scale data, lag behind specialized SER models: MERaLiON-2-10B achieves mid-tier performance (43--60\%), while SeaLLMs-Audio-7B remains the lowest-performing system. 
Closed-source LLMs perform competitively with merged segments but still underperform on fine-grained 7-class tasks, underscoring the need for task-specific acoustic representation learning.

\paragraph{Language-specific observations:}
MERaLiON-SER-v1 dominates in English, Chinese, and Tamil, while emotion2vec-seed slightly surpasses it for Malay. However, in case of primary 4-class evaluation, MERaLiON-SER-v1 achieves competitive performance as emotion2vec for Malay language.

\begin{figure}[h]
    \centering
    \includegraphics[width=0.8\linewidth]{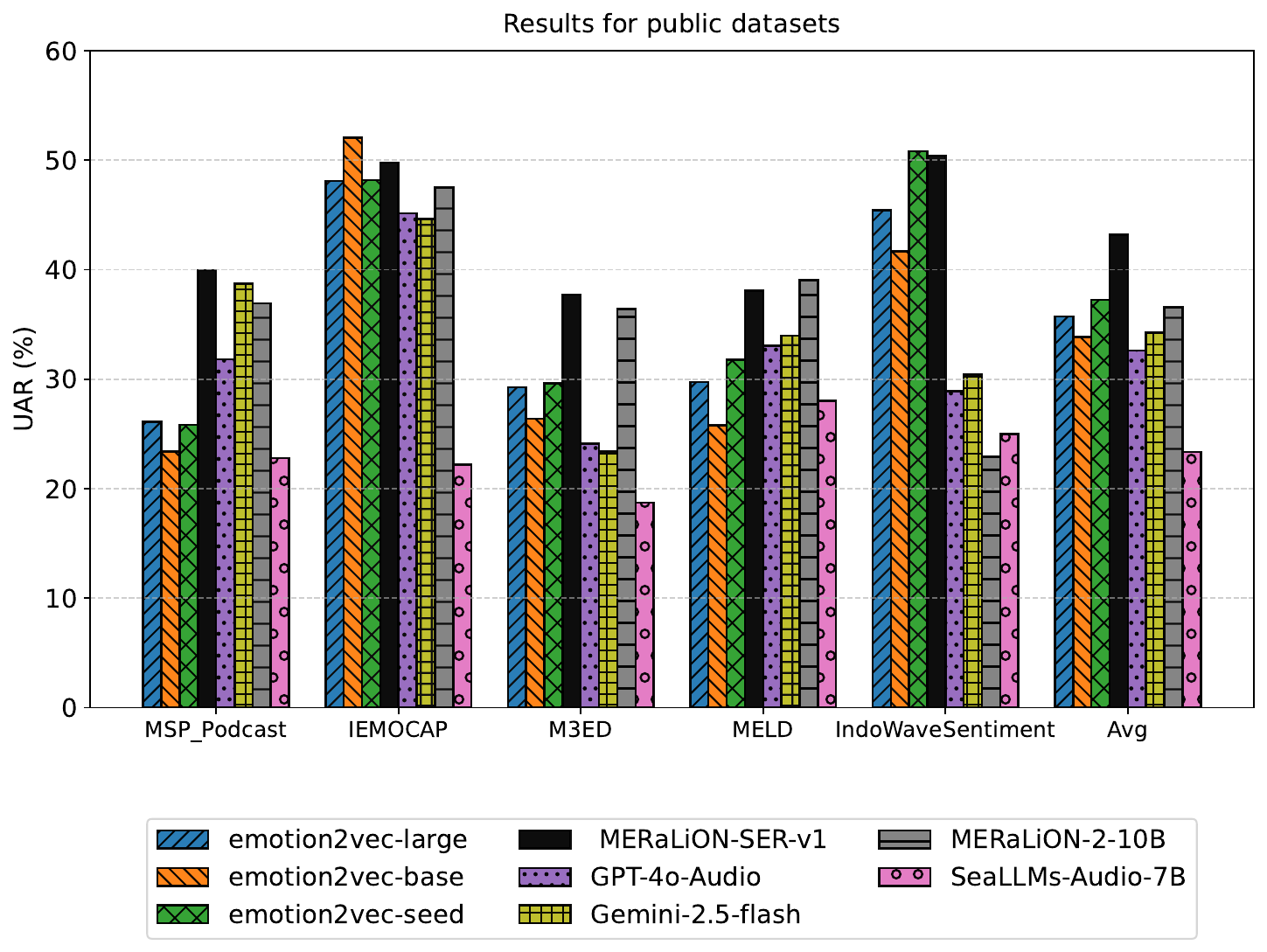}
    \caption{Results for public datasets with seven classes.}
    \label{fig:publicEval}
\end{figure}
\subsection{Evaluation on Public Multilingual Datasets}

We further evaluated the models on five publicly available emotion datasets: three English corpora (\textbf{MSP-Podcast}, \textbf{IEMOCAP}, \textbf{MELD}), one Chinese corpus (\textbf{M3ED}), and one Indonesian corpus (\textbf{IndoWaveSentiment}). 
These datasets differ in style—acted vs. spontaneous speech, monologue vs. dialogue, and high- vs. low-resource settings—allowing a robust test of cross-domain generalization. The results are shown  in Figure \ref{fig:publicEval} for different models.

\paragraph{Overall comparison:}
Across all datasets, \textbf{MERaLiON-SER-v1} again delivers the highest UAR, achieving 64--70\% on English corpora and around 57--60\% on Chinese and Indonesian data.
The emotion2vec has three variants and results shows that performance is not consistent across datasets and hence there is no single emotion2vec model that is better across all datasets.
The best open-source baseline (\textit{emotion2vec-seed}) lags by 4--6\% absolute UAR, while multimodal LLMs (e.g., GPT-4o-Audio, Gemini-2.5-flash) trail by 8--12\%. 
Despite having larger parameter counts, MERaLiON-2-10B remains below MERaLiON-SER-v1, indicating that model scale does not substitute for paralinguistic specialization.
\section{Related Work}
\subsection{Self-Supervised and Representation Learning for SER}  
Recent years have seen major advances via self-supervised learning (SSL) based speech models (e.g., Wav2Vec 2.0, HuBERT, WavLM) that learn contextualised acoustic features from large unlabeled corpora \citep{SSLreview}. These representations improve downstream SER—especially when fine-tuned with an emotion-specific dataset \citep{10447678}. There are several benchmarks that show the importance of speech SSL and large scale models like Whisper for speech emotion recognition \citep{emobox}, \citep{osman24_interspeech}. There are very limited attempts to create a large scale speech emotion model that is open source and can perform well on different languages. One notable attempt is emotion2vec where authors released open source models with different model sizes and training configurations \citep{emotion2vec}. Other open source model releases include models specifically trained by single dataset that is MSP podcast \citep{feng2025vox}.

\subsection{Audio-Language Models and Emotion Reasoning}  
The frontier of audio-text understanding has expanded significantly with the rise of audio large language models (AudioLLM) and multimodal agents (e.g., GPT-4o-Audio, Gemini). Past efforts to incorporate paralinguistic understanding into these models generally fall into three categories: (1) fine-tuning on specific emotional datasets \citep{Lin24,  kimparalinguistics, kang2024frozen, wang24blsp},, (2) knowledge distillation from specialized emotion recognition systems \citep{desta, desta2, wang24blsp}, and (3) translating acoustic cues directly into text prompts \cite{wu-etal-2025-beyond,CLAP4Emo,xu2024secap,wu2024empower}. Crucially, however, these general-purpose AudioLLM consistently prioritize speech semantics (transcription) over subtle paralinguistic cues (such as emotion and other speaker traits). As a result, their capacity for fine-grained emotion reasoning remains fundamentally limited. This persistent gap motivates the need for specialized SER models built explicitly for deep emotional representation.
In summary, while large pre-trained and multimodal models offer broad applicability, specialized SER models continue to play a vital role in capturing affective nuance. 
\section{Limitations and Future Work}
MERaLiON-SER demonstrates strong multilingual generalization and outperforms large multimodal models on speech emotion recognition task. Here, we highlight few limitations as well.
First, the model was trained on pseudo-labeled data for certain languages, which despite careful filtering may introduce label noise and bias. We conducted human-verified annotation task for evaluation set only. Our future work is to enhance this by adding small amount of manually labeled dataset for languages that does not have training dataset.
Second, this model provides seven emotion classes and hence can not be used for other categories such as contempt or for cases where complex emotional contexts is required such as sarcasm detection. However, dimensional scores make it possible to give more generalised emotion representation space.

Beyond these limitations, a promising research direction lies in integrating MERaLiON-SER within agentic frameworks, where emotion understanding can guide reasoning, dialogue planning, and empathetic response generation. Since existing Audio-LLMs are often limited in affective reasoning due to text-dominant alignment, a hybrid system where MERaLiON-SER acts as an emotional perception module could provide emotionally grounded signals to autonomous agents or conversational systems. Such extension would pave the way for next-generation empathetic Audio-LLMs and agentic AI systems capable of reasoning not only about emotions but also through them.
\section{Conclusions}
We released MERaLiON-SER, a robust speech emotion recognition model developed for English and Southeast Asian languages. 
In both regional (Singapore) and public multilingual benchmarks, MERaLiON-SER achieves superior and consistent performance across languages, emotion granularity, and segment durations. 
The results emphasize that, despite advances in multimodal LLMs, speech-only encoders with targeted paralinguistic learning remain essential for fine-grained affective reasoning and cross-lingual robustness.
\begin{ack}
The computational work for this article was fully performed on resources of the
National Supercomputing Centre (NSCC), Singapore (https://www.nscc.sg). We acknowledge Frank and his team from Asiastar International Consultancy Pte Ltd, who were engaged as an external consultancy to perform human annotations for the speech emotion annotation task. We also thank student intern Arjun Srinivas to help for data and automated emotion labeling work. This research is supported by the National Research Foundation, Singapore under its National Large Language Models Funding Initiative. Any opinions, findings, conclusions, or recommendations expressed in this material are those of the author(s) and do not reflect the views of the National Research Foundation, Singapore.
\end{ack}
\bibliography{custom}

\section{MERaLiON Team (alphabetical order)}
\label{sec:team}
Aw Ai Ti, Chen Fang Yih Nancy, Chiu Ying Lay, 
Ding Yang,
He Yingxu,
Jiang Ridong,
Li Jingtao,
Liao Jingyi,
Liu Zhuohan,
Lu Yanfeng,
Ma Yi,
Manas Gupta,
Muhammad Huzaifah Bin Md Shahrin,
Nabilah Binte Md Johan,
Nattadaporn Lertcheva,
Pan Chunlei,
Pham Minh Duc,
Sailor Hardik Bhupendra,
Siti Maryam Binte Ahmad Subaidi,
Siti Umairah Binte Mohammad Salleh,
Sun Shuo,
Tarun Kumar Vangani,
Wang Qiongqiong,
Won Cheng Yi Lewis,
Wong Heng Meng Jeremy,
Wu Jinyang,
Zhang Huayun,
Zhang Longyin,
Zou Xunlong










\end{document}